\documentclass[aps,prl,twocolumn,showpacs,amsmath]{revtex4}
\usepackage{graphicx}

\begin{document}

\title{Multi-phonon scattering and Ti-induced hydrogen dynamics
in sodium alanate }

\author{Jorge \'I\~niguez$^{1,2}$, T. Yildirim$^1$, T.
J. Udovic$^1$, E. H. Majzoub$^3$, M. Sulic$^4$, and C.
M. Jensen$^4$}

\affiliation{$^{1}$NIST Center for Neutron Research, National
Institute of Standards and Technology, Gaithersburg, MD 20899\\
$^{2}$Dept. of Materials Science and Engineering, University of
Maryland, College Park, MD 20742\\ $^{3}$Sandia National Laboratories,
7011 East Ave., MS 9403, Livermore, CA 94550\\ $^{4}$Department of
Chemistry, University of Hawaii, Honolulu, HI 96822}

\begin{abstract}
We use {\sl ab initio} methods and neutron inelastic scattering (NIS)
to study the structure, energetics, and dynamics of pure and Ti-doped
sodium alanate (NaAlH$_4$), focusing on the possibility of
substitutional Ti doping. The NIS spectrum is found to exhibit
surprisingly strong and sharp two-phonon features. The calculations
reveal that substitutional Ti doping is energetically possible. Ti
prefers to substitute for Na and is a powerful hydrogen attractor that
facilitates multiple Al--H bond breaking. Our results hint at new ways
of improving the hydrogen dynamics and storage capacity of the
alanates.
\end{abstract}

 \maketitle

Developing safe, cost-effective, and practical means of storing
hydrogen is crucial for the advancement of hydrogen and fuel cell
technologies.  Presently, there are three generic routes for the
solid-state storage of hydrogen: (i) physisorption as in many porous
carbon and zeolite materials, (ii) chemisorption as in metal hydrides,
and (iii) chemical reaction such as in complex metal hydrides. Among
the type (iii) materials, sodium alanate (NaAlH$_4$) has received
considerable attention because of its high hydrogen weight capacity
and low cost. The release of hydrogen from NaAlH$_4$ occurs via a
two-step reaction:
\begin{eqnarray}
\mbox{NaAlH$_4$} & \longleftrightarrow & \mbox{$\frac{1}{3}$
Na$_3$AlH$_6$ + $\frac{2}{3}$ Al + H$_2$ (3.7 wt\%)},
\nonumber\\
\mbox{Na$_3$AlH$_6$} & \longleftrightarrow & \mbox{3 NaH +
  Al + $\frac{3}{2}$ H$_2$ (1.8 wt\%)},
\label{eq:reactions}
\end{eqnarray}
yielding a total of 5.5~wt\% hydrogen. It was recently reported that a
few percent of Ti doping in NaAlH$_4$ renders accelerated and
reversible hydrogen release under moderate conditions~\cite{bog97}.
In spite of the extensive investigations of Ti-doped NaAlH$_4$ that
have resulted, little is known about the mechanism by which Ti
enhances the cycling kinetics of hydrogen~\cite{gro02a,kiy03}.  In
fact, even the location of the Ti atoms remains unclear. While it is
widely believed that they reside on the surface of the
material~\cite{gro00}, the possibility that Ti is substituted for Na
has also been suggested~\cite{sun02}, but convincing experimental or
theoretical evidence is still lacking.

Here we report neutron inelastic scattering (NIS) measurements of
phonon density of states and first-principles total energy and
dynamics calculations of pure and Ti-doped sodium alanates. We succeed
in characterizing the main features in the observed spectra, which
display surprisingly strong two-phonon contributions. Furthermore, the
calculations show that it is most energetically favorable for Ti to
substitute for Na, breaking several Al--H bonds in its vicinity. We
also find that Ti-doped NaAlH$_4$ can accommodate extra hydrogen near
the dopant. In addition, we examine the effect of the Ti dopants on
the vibrational spectrum of neighboring AlH$_4$ groups, which could
allow probing the Ti location in doped samples using high resolution
spectroscopic techniques.

Three powder samples were investigated: pure NaAlH$_4$, 2\% Ti-doped
NaAlH$_4$, and Na$_3$AlH$_6$.  NaAlH$_4$ was prepared as described in
Ref.~\onlinecite{san02}.  NaAlH$_4$ was doped with 2 mol percent
TiF$_{3}$ through mechanical milling according to standard
procedure~\cite{zid99}. Na$_3$AlH$_6$ was synthesized and purified by
the method of Huot {\it et~al.}~\cite{huo99} The NIS measurements
were performed using the Filter Analyzer Neutron Spectrometer (FANS)
at the NIST Center for Neutron Research~\cite{cop95} under conditions
that provided energy resolutions of 2--4.5\% in the range probed.

The calculations were performed within the plane-wave implementation
of the generalized gradient approximation (GGA-PBE~\cite{per96}) to
density functional theory (DFT) in the {\sf ABINIT}
package~\cite{gon02}. We used Troullier-Martins
pseudopotentials~\cite{tro93,fuc99} treating the following electronic
states as valence: 3$d$ and 4$s$ of Ti, 3$s$ and 3$p$ of Al, 3$s$ of
Na, and 1$s$ of H. We carefully tested the convergence of our
calculations with respect to the plane-wave cutoff and k-point mesh.
For example, for the Ti-doped alanate supercells (containing 96 atoms)
we used a cutoff of 1250~eV and a 3$\times$3$\times$2 k-point mesh.
The phonon spectrum of the pure compounds was computed using density
functional perturbation theory (DFPT)~\cite{bar01} as implemented in
{\sf ABINIT}. We calculated the phonons corresponding to two $2\times
2\times 2$ q-point grids. An interpolation scheme was then used so
that the powder-averaged NIS phonon spectra were computed for a
$4\times 4\times 4$ q-point grid within the incoherent
approximation~\cite{squ}.

For hydrogen dynamics studies, we first optimized the tetragonal
NaAlH$_4$ and monoclinic Na$_3$AlH$_6$ structures. The results,
summarized in Table~\ref{tab:struc}, agree reasonably well with
previous X-ray~\cite{bel82} and neutron~\cite{hau03,ozo} studies.  For
example, the reported NaAlH$_4$ lattice parameters (at
8~K)~\cite{hau03} are 4.98 and 11.15~\AA, respectively, while we
obtain 4.98 and 11.05~\AA. The reported H position (0.237,0.384,0.547)
is in good agreement with our result.

\begin{table}[b!]
\caption{Calculated structural parameters of NaAlH$_4$ and
Na$_3$AlH$_6$.}
\smallskip
\begin{center}
\begin{tabular}{cc@{\extracolsep{6mm}}ccc}
\hline\hline 
\multicolumn{5}{l}{NaAlH$_4$\; I4$_1$/a\;\;\; $a=4.98$ \AA\;, 
$c=11.05$ \AA} \\
\hline
Atom & Wyc. & $x$ & $y$ & $z$ \\
Al   & 4b   & 0 & 1/4 & 5/8 \\
Na   & 4a   & 0 & 1/4 & 1/8 \\
H    & 16f  & 0.2355 & 0.3918 & 0.5439 \\
\hline\hline
\multicolumn{5}{l}{Na$_3$AlH$_6$\; P2$_1$/n} \\
\multicolumn{5}{c}{$a=5.33$ \AA\;, $b=5.53$ \AA\;,
$c=7.68$\;, $\beta=$90.103$^{\circ}$} \\
\hline
Atom & Wyc. & $x$ & $y$ & $z$ \\
Al   & 2a   & 0 & 0 & 0 \\
Na   & 2b   & 0 & 0 & 1/2 \\
Na   & 4e   & 0.9897 & 0.4532 & 0.2535 \\
H    & 4e   & 0.1000 & 0.0481 & 0.2164 \\
H    & 4e   & 0.2281 & 0.3307 & 0.5437 \\
H    & 4e   & 0.1608 & 0.2673 & 0.9366 \\
\hline\hline
\end{tabular} 

\end{center}
\label{tab:struc}
\end{table}
\begin{figure}[t!]
\includegraphics[scale=0.6]{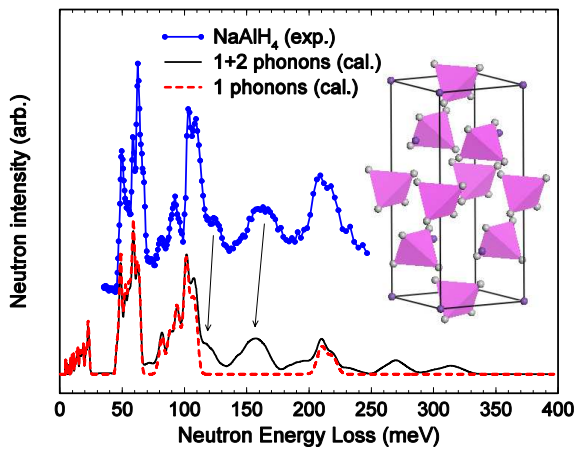}
\smallskip
\caption{Measured (top) and calculated (bottom) NIS spectra of
NaAlH$_4$.  The calculated 1 and 1+2 phonon contributions are
shown. The structure of NaAlH$_4$ is shown in the inset; grey
tetrahedra represent AlH$_4$ units.}
\label{fig:spec1}
\end{figure}
\begin{figure}[t!]
\includegraphics[scale=0.6]{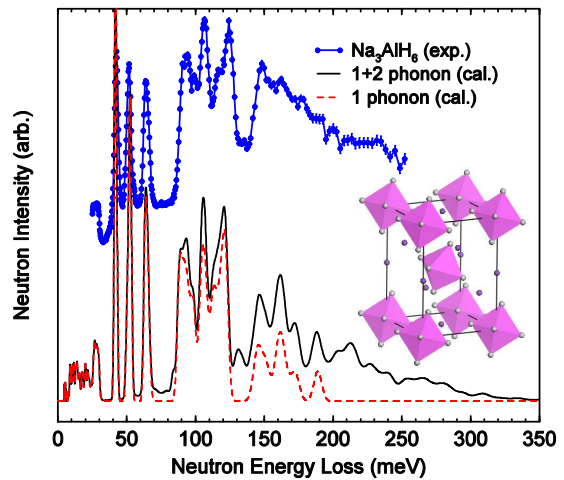}
\smallskip
\caption{Measured (top) and calculated (bottom) NIS spectra of
Na$_3$AlH$_6$. The calculated 1 and 1+2 phonon contributions are
shown. The structure of Na$_3$AlH$_6$ is shown in the inset; grey
octahedra represent AlH$_6$ units.}
\label{fig:spec2}
\end{figure} 

We also calculated the energy change associated with the two reactions
in Eq.~(\ref{eq:reactions}) and obtained, respectively, 24 and 42~kJ
per mol of H$_2$ released, compared to the experimental enthalpies of
reaction of 37 and 47~kJ/mol~\cite{bog00}. The agreement is reasonably
good taking into account (i) we cannot directly compare our energy
differences with the experimental enthalpies (because we do not
consider entropic contributions, {\it etc.}) and (ii) the error
associated with our first-principles approach is of the order of
5-10~kJ/mol. This error is estimated from the error in the calculated
cohesive energies of individual compounds; {\it e.g.}, we get $E_{\rm
coh}$=4.53~eV for the H$_2$ molecule, while the experimental value is
4.49~eV.

The vibrational spectra of NaAlH$_4$ and Na$_3$AlH$_6$ are shown in
Figs.~\ref{fig:spec1} and \ref{fig:spec2}, respectively.  The
calculated one-phonon spectrum does not reproduce several
experimentally observed features.  Yet, including two-phonon processes
yields results that are in excellent agreement with the observed
spectrum of both compounds. Such strong multi-phonon contributions are
unusual, but they seem to be typical of materials with AlH$_x$ groups.

\begin{figure}[t!]
\includegraphics[scale=0.6]{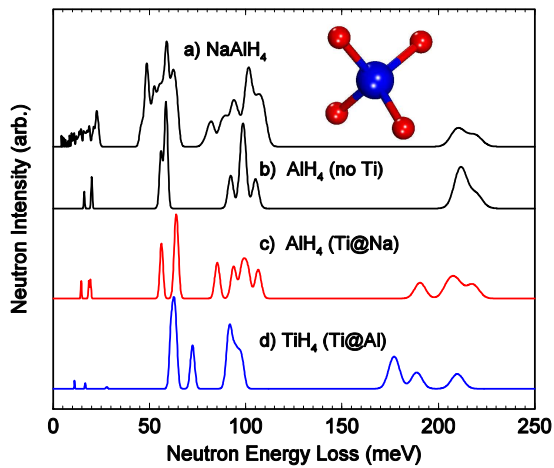}
\smallskip
\caption{Calculated one-phonon NIS spectra for (a) NaAlH$_4$ and for a
single MH$_{4}$ tetrahedron embedded in the lattice for various cases:
(b) pure alanate, (c) Ti$\rightarrow$Na, and (d) Ti$\rightarrow$Al.}
\label{fig:local}
\end{figure}

The nature of the different phonon bands in Fig.~\ref{fig:spec1} can
be determined by computing the modes of the individual AlH$_4$ groups
embedded in the crystalline matrix. (Note that the NIS spectrum is
dominated by hydrogen modes~\cite{squ}.)  As shown in
Fig.~\ref{fig:local}b, we obtain four distinct groups of modes that
correspond to the largest features in the one-phonon spectrum of
Fig.~\ref{fig:local}a. Inspection of the eigenvectors allows the
characterization of the modes. The lowest-frequency modes are AlH$_4$
rotations, while the peak above 200~meV consists of stretching modes
of the AlH$_4$ tetrahedron. The modes in the two intermediate sets are
a mixture of displacements and stretches.

Now we extend our calculations to investigate the possibility of
substitutional Ti doping of alanates. First we study whether Ti-doped
alanate is energetically stable and, if so, where the Ti dopant
goes. From the many substitutional and interstitial doping models that
could be tried, we choose two that are experimentally
motivated. Doping sodium alanate with solid TiCl$_3$ by dry
ball-milling results in formation of NaCl and partial desorption of
NaAlH$_4$, which leads to formation of aluminum
crystallites~\cite{gro02b}. Hence, it seems pertinent to study the
substitution of Al and Na by Ti. In the following we denote these
doping models by ``Ti$\rightarrow$Al'' and ``Ti$\rightarrow$Na''.

We consider supercells containing 16 NaAlH$_4$ formula units, and
substitute only one of the Al or Na atoms by Ti.  The cohesive
energies are obtained as the sum of the individual atom energies minus
the energy of the system. The results are given in
Table~\ref{tab:cohesive}; $E_{\rm coh}$ of the pure alanate has been
chosen as zero energy, so that positive entries indicate greater
stability than the pure system.

Note that the results in Table~\ref{tab:cohesive} give the energy
change in reactions of the form Ti~$+$~Na$_{16}$Al$_{16}$H$_{64}$
$\rightarrow$ Al~$+$~Na$_{16}$Al$_{15}$TiH$_{64}$, which involve
isolated atoms (Ti and Al in this case). These results thus measure
the relative stability of pure and doped systems. A positive (and
large) entry in the Table indicates that, in principle, it is feasible
to obtain the doped structure. Of course, in order to actually
accomplish this, it is necessary to identify appropriate reactants
that are able to release Ti easily and accommodate the atoms for which
Ti substitutes.

We find both Ti$\rightarrow$Na and Ti$\rightarrow$Al are energetically
more stable than pure alanate; {\sl i.e.} the system gains energy by
accepting a Ti dopant into the bulk and releasing a Na or Al atom. In
addition, Ti$\rightarrow$Na is found to be the most favorable
substitution.

\begin{table}[b!]
\caption{Calculated cohesive energies, given in eV and per 96-atom
supercell. The result for pure NaAlH$_4$ (231.922~eV) is taken as the
zero of energy. $E_{\rm coh}^{\rm atom}$ is the cohesive energy
obtained by allowing the atoms to relax but imposing the NaAlH$_4$
relaxed supercell; $E_{\rm coh}^{\rm atom}$(SP) is the same but
obtained from spin-polarized calculations; $E_{\rm coh}^{\rm full}$ is
the result obtained when allowing both atoms and cell to relax, and
$V$ the resulting volume of the 96-atom supercell in \AA$^3$.}
\label{tab:cohesive}
\smallskip
\begin{center}
\begin{tabular}{l@{\extracolsep{4mm}}rcrc}
\hline\hline
System & $E_{\rm coh}^{\rm atom}$ & $E_{\rm coh}^{\rm atom}$(SP) &
$E_{\rm coh}^{\rm full}$ & $V$ \\ 
\hline
Ti$\rightarrow$Al           &    0.075 &    0.408 &    0.113 & 1051.10 \\
Ti$\rightarrow$Na           &    0.911 &    1.192 &    1.024 & 1079.96 \\
Ti$\rightarrow$Na+Na$^v$    & $-$0.665 &    0.073 & $-$0.562 & 1059.52 \\
Ti$\rightarrow$Na+2Na$^v$   & $-$2.866 &          & $-$2.778 & 1059.27 \\
Ti$\rightarrow$Na(H)        &    1.316 &          &          &         \\
Ti$\rightarrow$Na+H         &    1.317 &          &          &         \\
\hline\hline
\end{tabular}\end{center}
\end{table}

The relaxed Ti$\rightarrow$Na structure presents H atoms that come
close to the Ti dopant. The shortest Ti--H distance is 2.05~\AA, to be
compared with the 2.39~\AA~Na--H distance in the pure
system. Consequently, the distance between the hydrogens close to the
dopant and their neighboring Al atoms is longer than the Al--H
distance in the pure system; we obtain 1.70 and 1.64~\AA,
respectively. This type of relaxation is to be expected, since Ti has
more electrons than Na to share with neighboring hydrogens.

It may seem surprising that Ti$\rightarrow$Na has a higher cohesive
energy than Ti$\rightarrow$Al. The typical valences of these atoms
certainly suggest otherwise. However, Ti seems to be relatively large
for the Al site. In the relaxed Ti$\rightarrow$Al structure we get a
Ti--H bond of 1.82~\AA, which is much longer than the 1.64~\AA~Al--H
bonds in pure sodium alanate. This size mismatch is the most likely
cause for Ti$\rightarrow$Al to be energetically less favorable.

The above results do not change significantly when the electrons are
allowed to spin polarize or when we allow both atoms and cell to relax
(see $E_{\rm coh}^{\rm atom}$(SP) and $E_{\rm coh}^{\rm full}$ columns
in Table~\ref{tab:cohesive}). The spin-polarized calculations predict
the Ti ion retains one unpaired electron in both Ti$\rightarrow$Na and
Ti$\rightarrow$Al.

A meaningful modification of the Ti$\rightarrow$Na doping model is to
introduce Na vacancies close to the Ti. Na vacancies should yield a
more balanced sum of valence charges and have been invoked in the
literature to argue that the Ti dopants reside in the bulk of the
system~\cite{sun02}. Our results for one and two Na vacancies are in
Table~\ref{tab:cohesive}, denoted by ``Ti$\rightarrow$Na+Na$^v$'' and
``Ti$\rightarrow$Na+2Na$^v$'', respectively. When spin polarization is
allowed, the Ti$\rightarrow$Na+Na$^v$ structure is found to be more
stable than the pure system, but significantly less stable than the
doping models considered above. On the other hand, the
Ti$\rightarrow$Na+2Na$^v$ structure is predicted to be quite unlikely.

The dynamics of the neighboring H atoms could be used as a local probe
for the Ti location.  However, dynamical calculations for the whole
96-atom supercell are very computationally demanding. Instead, we
calculated the vibrational spectrum of a single AlH$_4$ group near the
Ti dopant using the finite displacement technique~\cite{yil00}. The
results are shown in Fig.~\ref{fig:local}. In the Ti$\rightarrow$Na
case (panel~c), the Ti dopant mainly affects the high-frequency modes,
{\sl i.e.} those involving stretching of the AlH$_4$ tetrahedron. All
the modes in that group soften. The mode that softens most,
approximately from 210~meV to 190~meV, is dominated by the
displacement of H that is closest to the Ti, and essentially
corresponds to its oscillation along the Al--Ti direction. This
clearly indicates that the presence of Ti could facilitate breaking of
the Al--H bond.

In the Ti$\rightarrow$Al case (Fig.~\ref{fig:local}d), the dynamics
are modified quite differently since we deal with a TiH$_4$
group. This suggests that, by investigating the phonon spectrum of
Ti-doped NaAlH$_4$, one might determine whether Ti dopants go into the
bulk of the system and, if so, where they are located. Motivated by
this possibility, we measured the phonon spectrum of a 2\% Ti-doped
sample, but obtained a result essentially identical to that of pure
alanate shown in Fig.~\ref{fig:spec1}. However, it should be noted
that this does not rule out the possibility of substitutional doping
in our sample since the amount of Ti is very small, and thus any
dopant-induced feature in the spectrum should also be very small and
hard to distinguish from the noise. In addition, the NIS spectrum of
pure sodium alanate presents significant two-phonon intensity in the
175-200~meV energy range (see Fig.~\ref{fig:spec1}), which makes it
difficult to identify fine details. Higher resolution spectroscopic
measurements, such as Raman scattering, might help elucidate this
issue.

The above results suggest that Ti dopants may facilitate the breaking
of the Al--H bond. We explored this possibility by moving one H atom
to the immediate vicinity of the dopant and then relaxing the
system. The resulting structure, which we denote by
``Ti$\rightarrow$Na (H)'', is considerably more stable than the
original Ti$\rightarrow$Na doping model (see $E_{\rm coh}$ in
Table~\ref{tab:cohesive}). In fact, we found that it has not one but
two H atoms very close to the Ti.  The shortest Ti--H distance is
1.81~\AA, and the corresponding Al--H distance is 1.89~\AA, {\sl i.e.}
0.25~\AA~longer than in pure NaAlH$_4$.  Thus the Ti dopant can indeed
induce Al--H bond breaking, a necessary step for H$_2$ release.

\begin{figure}
\includegraphics[scale=0.45]{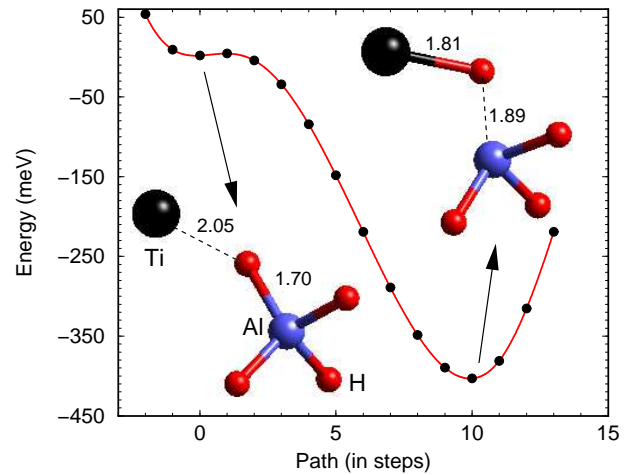}
\smallskip
\caption{Energy along the path from Ti$\rightarrow$Na to
Ti$\rightarrow$Na~(H) structures (see text). Insets show
the local structure and bond distances at the two minima. 
The zero of energy is arbitrary.}
\label{fig:path1}
\end{figure}

These results indicate that the Ti$\rightarrow$Na structure is a local
minimum. In Fig.~\ref{fig:path1} we show the energy change along the
transition path leading from Ti$\rightarrow$Na to
Ti$\rightarrow$Na~(H). The Ti$\rightarrow$Na minimum is very
shallow. The calculated energy barrier is around 2.5~meV$\approx$30~K,
indicating that Al--H bonds would immediately break in the presence of
Ti. Associated with the shallow well is a collective mode of the
Ti$\rightarrow$Na structure whose frequency can be roughly estimated
to be $\sim$80~meV.  Yet, the frequency of the embedded AlH$_4$ mode
that we related to the Al--H bond breaking is about 190~meV (see the
discussion of Fig.~\ref{fig:local}c). The reduction from 190 to 80~meV
is related to other atomic rearrangements, {\sl e.g.}, the second H
coming close to the Ti atom, displacements of the AlH$_4$ groups, {\sl
etc}.

Along these lines, we also considered a less obvious possibility,
namely, that Ti drags extra hydrogens into the
system. Table~\ref{tab:cohesive} shows the result for
``Ti$\rightarrow$Na+H'', which corresponds to placing one extra H in
the vicinity of the Ti atom. (In the calculation of this cohesive
energy, the extra hydrogen was assigned one half of the energy of the
H$_2$ molecule, so that the resulting $E_{\rm coh}$ measures stability
against H$_2$ release.)  This structure turns out to be very
stable. We find a Ti--H bonding distance of 1.82~\AA $\;$ and several
AlH$_4$ groups approaching the Ti dopant.

In conclusion, we have used first-principles methods and neutron
inelastic scattering to study pure and Ti-doped sodium alanate
(NaAlH$_4$), a material that holds great promise for reversible
hydrogen storage.  The total energy calculations indicate that
substitutional Ti doping in NaAlH$_4$ is energetically stable. We find
that the dopant prefers to substitute for Na and attracts several
hydrogen atoms, softening and breaking the corresponding Al--H
bonds. We also find it energetically favorable for the Ti to drag
extra H atoms into the system. These results point to an interesting
direction for future research, namely, the possibility of producing a
new material, sodium-titanium alanate, that might benefit from the
ability of Ti to accommodate extra hydrogens in its vicinity and thus
exhibit improved H-storage capabilities.

\acknowledgments

We acknowledge fruitful discussions with D. A. Neumann and Mei-Yin
Chou.

\end{document}